\documentclass[aip,cha,reprint]{revtex4-1}
\usepackage{mathtools}
\usepackage{amsmath}
\usepackage{amssymb}
\usepackage{graphicx}
\usepackage{dcolumn}
\usepackage{bm}
\usepackage{natbib}		
\usepackage{color}

\newcommand {\e} {\varepsilon}

\begin{document}
\title{Deterministic active particles in the overactive limit}

\author{Arkady Pikovsky}
\affiliation{Institute of Physics and Astronomy, University of Potsdam,
Karl-Liebknecht-Str. 24/25, 14476 Potsdam-Golm, Germany}
\email{pikovsky@uni-potsdam.de}

\date{\today}

\begin{abstract}
We consider two models of deterministic active particles in an external potential. In the limit where the speed of a particle is fixed, both models coincide and can be formulated as a Hamiltonian system, but only if the potential is time-independent. If the particles are identical, their interaction via a potential force leads to conservative dynamics with a conserved phase volume. In contrast, the phase volume is shown to shrink for non-identical particles. 
\end{abstract}

\maketitle

\begin{quotation}
Active particles possess an internal energy source, which maintains motion with a preferable speed. Applications of this concept include living organisms (from cells to birds) but also many engineered devices like robots and drones. If the ``speed control'' is perfect, an active particle moves with a constant speed, and the external forces (from the environment or from other particles) can only the direction of the velocity. We focus on this overactive limit and demonstrate that the dynamics of an overactive particle in a static external potential is Hamiltonian. However, if the potential is time-dependent, or there is an interaction of nonidentical particles, the dynamics is no more Hamiltonian but dissipative.
\end{quotation}

\section{Introduction}
\label{sec:intro}
Active, or self-propelled particles, are elementary models of the dynamics out of equilibrium. Their study is relevant not only for living systems but also for
many setups in physics and engineering, where there is an energy flux supporting a directed motion of the agents. Large ensembles of active particles constitute active matter, showing many interesting patterns of behavior~\cite{chate2020dry,gompper20202020,aranson2013active}. However, interesting effects are also observed within a small number of particles placed in a complex environment~\cite{bechinger2016active,peruani2018cold}. In many cases, noise is essential in the dynamics, and one speaks on Brownian active particles~\cite{erdmann2000brownian,romanczuk2012active}. 

This paper focuses on a less explored case of purely deterministic active particles.
Such a noise-free case is probably, less relevant for living systems but can be realized with engineered objects like robots and drones.
We will show that even the dynamics of one active particle in an external potential is non-trivial. 

In the literature, there are two basic models of deterministic active dynamics;
both are based on adding active forces to the standard mechanical equations of motion.
We will introduce these models in Section~\ref{sec:tm}. Remarkably, both models beget the same limit if the activity that fixes a particle's speed becomes very strong. We refer to this situation as the \textit{overactive} limit and explore it in detail in Section~\ref{sec:ham}. Noteworthy,
a single overactive particle in a static external potential is a Hamiltonian system~\cite{Aranson-Pikovsky-22}. However, the Hamilton function is so peculiar that one cannot extend the Hamiltonian description to a non-stationary potential and to the interaction of several particles. We show that, under certain conditions, in these cases, the dynamics becomes dissipative (Section~\ref{sec:cdd}). We conclude the paper by discussing possible extensions in Section~\ref{sec:concl}.


\section{Two models for activity and their overactive limits}
\label{sec:tm}
This section will present two models for active deterministic particles from the literature 
and demonstrate that they have the same form in the overactive limit. In this paper, we 
consider the simplest two-dimensional setup. 

\subsection{Self-propelled particles}
We consider a two-dimensional motion of a particle with mass $M$ under generic external force $\vec{f}$ that obeys equations
\begin{align}
\dot{\vec{r}}&=\vec{v}\;,\label{eq:spp1}\\
M\dot{\vec{v}}&=\e^{-1}(V^2-v^2)\vec{v}+\vec{f}(\vec{r},t)\;,\label{eq:spp2}
\end{align}
and call this case a ``self-propelled particle''. 
Here, an ``internal active force'' is introduced~\cite{erdmann2000brownian,erdmann2002excitation,erdmann2005attractors}, 
which acts to keep the speed of 
the particle as close as possible to the preferred speed $V$. 
Parameter $\epsilon$ defines the
rate at which this preferred speed is settled. Note that for $\e=\infty$, the active force vanishes, and the particle obeys the second Newton's law.
One can see that a steady state $\vec{v}=0$ is possible at the positions where the external force $\vec{f}$ vanishes, 
but is always unstable because, at small velocities, the active force corresponds to negative linear friction.

It is convenient to introduce the speed $v(t)=|\vec{v}|$ and the 
velocity direction (unit vector) $\vec{n}(t)=\vec{v}/|\vec{v}|$ of the particle. 
Then equations for these variables follow from \eqref{eq:spp2}:
\begin{align}
\dot v&=\frac{1}{M\e}(V^2-v^2)v+\frac{\vec{f}\vec{n}}{M}\;, \label{eq:sppvab}\\
\dot{\vec{n}}&=
\frac{\vec{f}-\vec{n}(\vec{f}\cdot\vec{n})}{Mv}\;.\label{eq:sppvdir}
\end{align}

Below we will focus on the case of strong activity, i.e., on the case where $\e$ is small.
Then the equations can be simplified by assuming that the fast variable $v$ is enslaved
by other two variables $\vec{r},\vec{n}$. If we write $v=V+\e v_1(\vec{r},\vec{n})$ and substitute this in
\eqref{eq:sppvab}, we obtain in order $\e^0$
\[
v_1(\vec{r},\vec{n})=\frac{\vec{f}\vec{n}}{2V^2}\;.
\]
Substitution of this to \eqref{eq:sppvdir} yields
\begin{equation}
\dot{\vec{n}}=
\frac{\vec{f}-\vec{n}(\vec{f}\cdot\vec{n})}{MV(1+\e \vec{f}{\vec{n}(2V^3)^{-1}})}\approx
\frac{\vec{f}-\vec{n}(\vec{f}\cdot\vec{n})}{MV}(1-\e \frac{\vec{f}\vec{n}}{2V^3}).
\label{eq:sppappr}
\end{equation}
where in the last transformation, we keep the leading order in $\e$.

Two equations \eqref{eq:spp1} and \eqref{eq:sppappr} describe the dynamics for of a self-propelled particle
for a strong activity.
This result allows for going to an \textit{overactive limit} $\e\to 0$, resulting in the system
\begin{equation}
\begin{aligned}
\dot{\vec{r}}&=V\vec{n}\;,\\
\dot{\vec{n}}&=\frac{\vec{f}-\vec{n}(\vec{f}\cdot\vec{n})}{MV}\;.
\end{aligned}
\label{eq:oal}
\end{equation}
 
\subsection{Overdamped particles}
In the literature, also another model for deterministic active dynamics has been suggested: 
\begin{gather*}
\dot{\vec{r}}=\vec{v}\;,\\
M\dot{\vec{v}}=-\gamma \vec{v}+\vec{F}+\vec{f}(\vec{r},t)\;.
\end{gather*}
Together with the external force $\vec{f}$ as above, one introduces linear friction $\sim\gamma$ and an internal active force $\vec{F}$.
Notice, that the  self-propelled case above corresponds to choosing $\vec{F}=g(v)\vec{v}$ (i.e., the internal force is directed along the
velocity and depends on it only), so one can combine friction and $\vec{F}$ in one velocity-dependent term like in \eqref{eq:spp2}.

However, one supposes that $\vec{F}$ does not depend on the velocity explicitly. Furthermore, one takes the \textit{overdamped}
limit $M\to 0$. In this limit, $\vec{v}$ is the fast variable, and it relaxes to the slow manifold, on which it is 
enslaved by forces $\vec{F}$ and $\vec{f}$:
\[
\dot{\vec{r}}=\vec{v}=\gamma^{-1}(\vec{F}+\vec{f})
\]

The next step is to specify the active force $\vec{F}$. One assumes that there is an intrinsic unit vector $\vec{m}$ which governs
the direction of the active force and the amplitude of the force is constant
\[
\vec{F}=F_0\vec{m}\;.
\]
The next step is to write an equation for $\vec{m}$. In Refs.~\onlinecite{Lam_2015,PhysRevLett.122.068002,baconnier2022selective,Damascena_etal-22} it was assumed that this vector 
rotates toward the direction of the velocity:
\[
\dot{\vec{m}}=\frac{1}{\epsilon} (\vec{m}\times \dot{\vec{r}})\times \vec{m}\;.
\]
Substituting here $\dot{\vec{r}}=\gamma^{-1}(\vec{F}+\vec{f})$, one gets
\[
\dot{\vec{m}}=(\epsilon\gamma)^{-1}(\vec{f}-\vec{m}(\vec{f}\cdot\vec{m}))\;.
\]
Summarizing one has a system
\begin{align}
\gamma\dot{\vec{r}}&=F_0\vec{m}+\vec{f}\;,\label{eq:odlr}\\
\gamma\dot{\vec{m}}&=\epsilon^{-1}(\vec{f}-\vec{m}(\vec{f}\cdot\vec{m}))\;.\label{eq:odlm}
\end{align}

These equations are very similar to \eqref{eq:oal} but with an additional force term in \eqref{eq:odlr}.

Let us renormalize time by  $\gamma\epsilon$, and denote $\epsilon F_0=V$. Then the equations of motion read
\begin{equation}
\begin{aligned}
\dot{\vec{r}}&=V\vec{m}+\epsilon \vec{f}\;,\\
\dot{\vec{m}}&=(\vec{f}-\vec{m}(\vec{f}\cdot\vec{m}))\;.
\end{aligned}
\label{eq:odl2}
\end{equation}
In the limit $\epsilon\to 0$, the system \eqref{eq:odl2} almost exactly reduces to \eqref{eq:oal}.
The difference is that $\vec{n}$ in \eqref{eq:oal} is the direction of velocity, while $\vec{m}$
in \eqref{eq:odl2} is an ``internal'' unit vector determining the direction of the active force.
We note that the rescaling of time and of the preferred speed include $\epsilon$; this means that
the limit $\epsilon\to 0$ should be taken together with limits $\gamma\sim \epsilon^{-1}$, $F\sim \epsilon^{-1}$.

\section{Hamiltonian dynamics of the overactive particles}
\label{sec:ham}
\subsection{Hamiltonian formulation of the equations of motion}
We have seen that in a limit where the speed of an active particle is fixed,
both models lead to the same equations \eqref{eq:oal}. Below we focus on the case
where the external force has potential $\vec{f}=-\nabla u(\vec{r})$. Then the basic equations 
in the overactive limit are
\begin{align}
\dot{\vec{r}}&=V\vec{n}\;,\\
\dot{\vec{n}}&=\frac{-\nabla u+\vec{n}(\nabla u\cdot\vec{n})}{MV}\;.
\end{align}
Let us show that these equations can be written as a Hamiltonian system with the Hamiltonian
\[
H(\vec{p},\vec{r})=V\lvert\vec{p}\lvert-\exp[-\frac{u(\vec{r})}{MV^2}]=0\;.
\]
The Hamiltonian equations of motion read
\begin{align}
\dot{\vec{r}}&=\frac{\partial H}{\partial \vec{p}}=V\frac{\vec{p}}{\lvert \vec{p}\lvert}\;,\\
\dot{\vec{p}}&=-\frac{\partial H}{\partial \vec{x}}=-\frac{1}{MV^2}\exp[-\frac{u}{MV^2}]\nabla u\;.
\end{align}
We now take into account that because the Hamilton function vanishes, $\lvert \vec{p}\lvert=V^{-1}\exp[-\frac{u}{MV^2}]$,
so that we can introduce the unit vector $\vec{n}$ according to $\vec{n}=V\exp[\frac{u}{MV^2}]\vec{p}$. A straightforward calculation of the derivative
of this vector yields
$
\dot{\vec{n}}=
(MV)^{-1}(\nabla u\cdot\vec{n})\vec{n}-(MV)^{-1}\nabla u
$. 
Thus, the resulting system is exactly \eqref{eq:oal}.

It is convenient to introduce an angle $\theta$ determining the direction of the velocity $\vec{n}=(\cos\theta,\sin\theta)$, so that the equations for the overactive particle
can be written as
\begin{equation}
\begin{aligned}
\dot x&=V\cos\theta\;,\\
\dot y&=V\sin\theta\;,\\
\dot\theta&=\frac{1}{MV}(-u_y\cos\theta+u_x\sin\theta)\;.
\end{aligned}
\label{eq:xyt}
\end{equation} 
We stress here that the variables $x,y,\theta$ are natural for numerical simulations, but they are not canonical ones.

Noteworthy, the Hamiltonian description of an overactive particle in a two-dimensional
potential is fully analogous to the Hamiltonian representation of the ray dynamics in optics (and in the other wave fields)~\cite{Kravtsov-Orlov-90}. In the latter case the Hamiltonian reads $H=|\vec{p}|-n(\vec{r})=0$, where $n(\vec{r})$ is the refraction index. We see that for the overactive particles, the effective refraction index is related to the potential as $n(\vec{r}) \sim \exp[-u(\vec{r})]$. We also mention another approach to the dynamics
of active particles, where an analogy with ray optics has been recently established~\cite{Ross_etal-22}.

\subsection{Case of small velocity}
It is instructive to consider the case of small velocity $V$, which leads to a separation of time scales. As one can easily see from \eqref{eq:xyt}, in this case, the coordinates $x,y$ are slow variables, and the angle $\theta$ is a fast variable. By introducing a direction $\beta$, opposite to the gradient of the potential, $u_x=-\cos\beta|\nabla u|$, $u_y=-\sin\beta |\nabla u|$, we can write the fast dynamics as~\cite{chepizhko2013diffusion,peruani2018cold}
\[
\dot\theta=\frac{|\nabla u|}{MV}\sin(\beta-\theta)\;.
\]
One can see that the fast motion redirects the particle's velocity toward the local steepest descent $\theta\to\beta$. After this, the particle slowly moves along the steepest descent toward a minimum of the potential. However,
close to the minimum, the value of $|\nabla u|$ becomes small, and the scale separation is no longer valid. Close to a minimum, a potential can be generally represented as a harmonic one. Therefore, studying the dynamics in a harmonic potential is especially relevant for slow particles.

\subsection{Motion in a harmonic potential}

\begin{figure}
	\centering
	\includegraphics[width=\columnwidth]{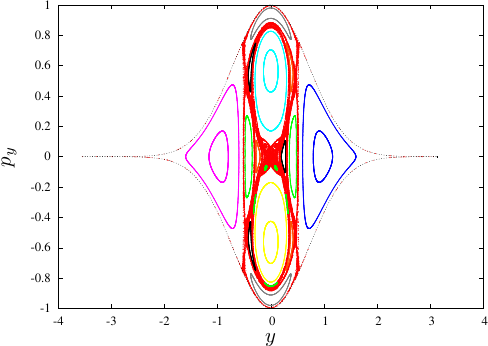}
	\caption{Poincare map for  $u=(x^2+2y^2)/2$. Poincar\'e section: $x=0$, $\dot x>0$. Other parameters: $M=V=1$. There is one chaotic domain (red dots) and domains with quasiperiodic dynamics (dots with different colors).}
	\label{fig:pmom00}
\end{figure}

Here we demonstrate that an overactive dynamics of a particle in a simple harmonic potential exhibits a typical for the Hamiltonian dynamics picture of a  divided phase space
with chaotic and quasiperiodic trajectories. (For the dissipative dynamics of deterministic active particles far from the overactive limit in a harmonic and other confining potentials, see Refs.~\onlinecite{erdmann2000brownian,erdmann2002excitation,erdmann2005attractors,PhysRevLett.122.068002,Damascena_etal-22}.) 
We demonstrate this with a two-dimensional Poincar\'e map
for a particle in a potential $u(x,y)=(x^2+2y^2)/2$, Fig.~\ref{fig:pmom00}. Here, although we solve equations in the natural coordinates $x,y,\theta$, we plot the canonical coordinates $y,p_y=V^{1}\sin\theta \exp[-y^2(MV^2)^{-1}]$ to ensure that the phase volume in the Poincar\'e section is conserved.  Note that the allowed domain of values of $p_y$ is bounded by a Gaussian curve, depicted in Fig.~\ref{fig:pmom00} with black dotted lines.

\subsection{Statistical validity of the overactive limit}
In our derivation of the overactive limit in Section~\ref{sec:tm}, we assumed that
the internal active force is much stronger than the external force acting on a particle. 
For a harmonic potential, the external force is unbounded, so the question arises if the overactive limit can be violated in the course of the dynamics. Such trajectories indeed correspond to the symmetry axes of the potential: $y=0, \theta=0,\pi$ and $x=0,\theta=\pi/2,3\pi/2$. The motion along these axes is one-dimensional, so a particle in the overactive limit moves straightforwardly with a constant velocity and enters the domains where
the external force becomes very large. In these domains, corrections to the overactive dynamics should be considered.

The simplest case is the overdamped dynamics Eqs.~\eqref{eq:odl2}. The one-dimensional dynamics along the $x$-axis in a potential $u=(ax^2+by^2)/2$ 
reduces to $\dot x= V-\epsilon ax$, with a stable steady state at $x=V(\epsilon a)^{-1}$.
In the case of a self-propelled particle (Eqs.~\eqref{eq:spp1},\eqref{eq:spp2}),
the one-dimensional dynamics along the $x$-axis, in the same potential, reduces to the Rayleigh equation $M\ddot x-\varepsilon^{-1}(V^2-v^2)v+ax=0$, which describes relaxation oscillations with the amplitude $x_{max}\sim \varepsilon^{-1}$. In both cases, the overactive limit becomes violated in the full equations, and the particle either stops or turns around at large distances from the minimum of the potential.

Remarkably, such events are practically not observed in the direct simulations of the two-dimensional dynamics. Indeed, as we argue below, the one-dimensional trajectories described above are transversally unstable, and their appearance is, therefore, extremely unprobable.  

To see this, we take a potential $u=(ax^2+by^2)/2$, so that the overactive dynamics \eqref{eq:xyt} reduces to
\begin{gather*}
\dot x=V\cos\theta,\quad\dot y=V\sin\theta,\\
\dot\theta=(-by\cos\theta+ax\sin\theta)(MV)^{-1}\;.
\end{gather*}

\begin{figure}[!htb]
	\centering
	\includegraphics[width=0.9\columnwidth]{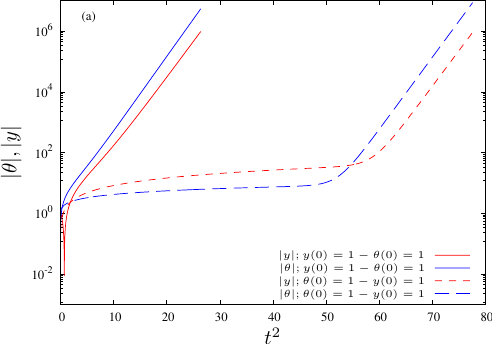}
	\includegraphics[width=0.9\columnwidth]{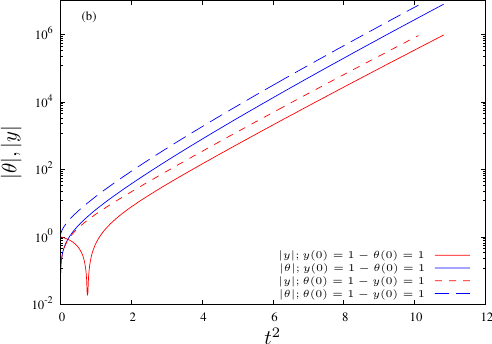}
	\caption{Evolution of solutions of Eqs.~\eqref{eq:erreq} with $M=V=1,\;x_0=1$. Left panel: $b=2,a=1$; right panel: $b=1,a=2$.
		We show $|y|,|\theta|$ as functions of $t^2$,
		for two different initial conditions $(y(0)=1,\;\theta(0)=0)$ and $(y(0)=0,\;\theta(0)=1)$. At large time the solutions follow the asymptotics  $\sim \exp[\text{const}\cdot t^2]$.}
	\label{fig:err}
\end{figure}

A trajectory $x^*(t)=Vt+x_0$, $\theta^*(t)=y^*(t)=0$ escapes to infinity. Let us consider a small transversal perturbation  $y,\theta$ in a vicinity of this trajectory:
\[
	\dot y=V\theta\;,\quad
	\dot\theta=\frac{1}{MV}(-by+a (V t+x_0)\theta)\;.
\]
This linear system can be reformulated as a single equation for $y$
\begin{equation}
	M\ddot y -(at+\frac{a x_0}{V}) \dot y+by=0\;.
	\label{eq:erreq}
\end{equation}
One cannot solve this linear equation explicitly, but for large $t$ if one keeps the dominating term $\sim at$ only, then
\[
\ddot y-a M^{-1} t\dot y=0
\]
has a solution 
\[
y=C_1\text{erfi}(\sqrt{\frac{a}{2}} t)+C_2,
\]
where $\text{erfi}$ is the imaginary error function which asymptotically grows as $\sim t^{-1} e^{at^2/2}$.
Thus one expects $y$ in the full equation also to grow with the same asymptotics. This is illustrated in Fig.~\ref{fig:err}, where we show that solutions of \eqref{eq:erreq} after a transient indeed grow $\sim \exp[\text{const}\cdot t^2]$.

\begin{figure}[!htb]
	\centering
	\includegraphics[width=0.8\columnwidth]{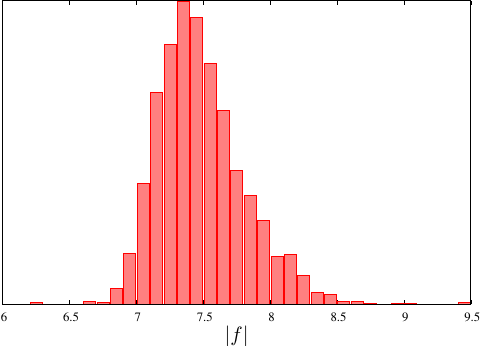}
	\caption{A histogram of the maximal values of the force in runs of duration $5\cdot 10^5$ ($M=V=1$).  }
	\label{fig:hist}
\end{figure}

This very strong transversal instability of the escaping solutions makes their appearance in numerical simulations of the overactive dynamics \eqref{eq:oal} extremely unprobable.
To illustrate this, we performed a statistical analysis of a maximal observed force $|f|=\sqrt{f_x^2+f_y^2}$ over a long (time interval $5\cdot 10^5$) chaotic trajectory of a particle in a harmonic potential $u(x,y)=(x^2+2y^2)/2$. 
 The histogram of these maximal values for 2000 runs is shown in Fig.~\ref{fig:hist}.
 In no run, a value exceeding $9.5$ has been observed. This indicates for a ``statistical validity'' of the overactive limit.

\section{Conservative and dissipative dynamics in time-dependent potentials and for interacting particles}
\label{sec:cdd}

Unfortunately, the Hamiltonian formulation of the dynamics of an overactive particle is restricted to a single particle in a time-independent potential. Already an inclusion of an explicit time dependence in the potential $u(x,y,t)$ does not allow to build Hamiltonian equations according to the derivation presented in Section~\ref{sec:ham} (because the value of the Hamilton function in this formulation is fixed, it cannot depend on time explicitly).
Therefore we present below an approach to characterize the conservativity/dissipativity of the dynamics based on the analysis of the phase volume divergence.

\subsection{Phase volume conservation}

According to the Liouville theorem, the phase volume of a Hamiltonian system is conserved instantaneously. This result is valid, however, only for the phase volume expressed in the canonical variables. For the same system expressed in other, non-canonical variables, the phase volume is conserved not instantaneously but on average in time. We illustrate this with the basic example above of a single particle in a time-independent potential. In the natural variables, the equations are written as a three-dimensional system  \eqref{eq:xyt}. It is straightforward to calculate the divergence rate $\alpha$ of the phase volume $W$:
\begin{equation}
	\begin{gathered}
	\alpha(t)=W^{-1}\frac{d W}{dt}=\partial_x\dot x+\partial_y\dot y+\partial_\theta \dot\theta=\\
	(MV^2)^{-1}(u_y\sin\theta+u_x\cos\theta)=\\=(MV^2)^{-1}(u_y\dot y+u_x\dot x)=(MV^2)^{-1}\frac{du}{dt}\;.
	\end{gathered}
\label{eq:phvol1}
\end{equation}
One can see that the instantaneous divergence of the phase volume fluctuates, but its time average over a long time interval vanishes for a statistically stationary regime:
\[
\langle \alpha(t)\rangle_T=\frac{1}{T}\int_0^T \alpha(t')dt'=\frac{u_T-u_0}{MV^2 T}\underset{T\to\infty}{\to} 0\;.
\]

We will adopt the phase volume conservation on average as a measure for the conservativity of the dynamics. Below we will check several situations where the Hamiltonian formulation is impossible for conservativity.

\subsection{Time-dependent potential}
\begin{figure}[!htb]
	\centering
	\includegraphics[width=0.45\textwidth]{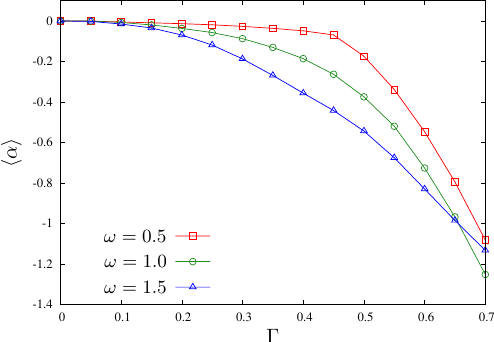}
	\caption{Dependence of the average divergence rate on the level of time-modulation for different $\omega$. Average time $T=10^6$. Parameters: $a=1,\;b=2,\;M=V=1$.}
	\label{fig:nstat}
\end{figure}

If the potential has an explicit time dependence $u(x,y,t)$, then relation \eqref{eq:phvol1} is modified as
\[
\alpha(t)=(MV^2)^{-1}\left(\frac{du}{dt}-\frac{\partial u}{\partial t}\right)
\]
and one cannot conclude that the time average of $\alpha$ vanishes. We performed numerical simulations for a particle in a potential 
\[
u(x,y,t)=\frac{a(1+\Gamma \cos(\omega t))x^2+ b(1+\Gamma\sin(\omega t))y^2}{2}\;.
\]
The average value of $\alpha$ vs. the level of time-modulation is shown in Fig.~\ref{fig:nstat}. One can see that $\langle \alpha(t)\rangle_T$ gradually decreases with $\Gamma$. This allows for a conclusion that the dynamics of an overactive particle motion in a time-dependent potential is dissipative.

\subsection{Interacting particles}

Above, we concentrated on the dynamics of an overactive particle in an external potential. Here we explore interaction between the particles, defined by a two-particle potential $U(\vec{r}_1,\vec{r}_2)$ (for numerical explorations of complex patterns appearing in populations of interacting active particles far from the overactive limit, see Refs.~\onlinecite{touma2010self,Alsayegh_etal-16,PhysRevLett.120.208001,baconnier2022selective}). For simplicity, we consider just two particles, so an extension for a population with pairwise interactions is obvious. The equations of motion for particles $1,2$ read
\begin{equation}
	\begin{aligned}
		\dot x_{1,2}&=V_{1,2}\cos\theta_{1,2}\;,\\
		\dot y_{1,2}&=V_{1,2}\sin\theta_{1,2}\;,\\
		\dot\theta_{1,2}&=\frac{1}{M_{1,2}V_{1,2}}(-\frac{\partial U}{\partial y_{1,2}}\cos\theta_{1,2}+\frac{\partial U}{\partial x_{1,2}}\sin\theta_{1,2})\;.
	\end{aligned}
	\label{eq:2xyt}
\end{equation} 
In general, parameters $M,V$ for two particles are different.
Similar to Eq.~\eqref{eq:phvol1}, we calculate the divergence of the total phase volume:
\begin{equation}
	\begin{gathered}
		\alpha(t)=\sum_{m=1,2} \left(\frac{\partial \dot x_m}{\partial x_m}+\frac{\partial \dot y_m}{\partial y_m}+\frac{\partial\dot\theta_m}{\partial \theta_m}\right) =\\
		=\sum_{m=1,2} (M_m V_m^2)^{-1}\left(\dot x_m\frac{\partial U}{\partial x_m}+
		\dot y_m\frac{\partial U}{\partial y_m}\right)\;.
	\end{gathered}
\label{eq:phvol2}
\end{equation}

One can see that the divergence reduces to a total derivative of the interaction potential only in the symmetric case where $M_1V_1^2=M_2V_2^2$: then, similarly to \eqref{eq:phvol1},
\[
\alpha(t)=(MV^2)^{-1}\frac{d U}{dt}\;.
\]
In this case, the long-time average of the divergence vanishes, and the dynamics is conservative. In the asymmetric case,  $M_1V_1^2\neq M_2V_2^2$, we cannot generally expect conservativity. 

To check this prediction, we considered a population of particles in a harmonic external potential $u(x,y)=(x^2+2y^2)/2$, with a repulsive two-particle interaction according to a potential 
\begin{equation}
U_{ij}(R)=\begin{cases}
	D|(R/\sigma)^2-1|^7& R<\sigma\;,\\
	0 &R\geq \sigma\;,
\end{cases}
\label{eq:intpot}
\end{equation}
with some constant $D$. Here $R=\sqrt{(x_i-x_j)^2+(y_i-y_j)^2}$ is the distance between particles, and parameter $\sigma$ determines a distance at which the repulsive force disappears. The force is calculated as
\[
\vec{f}_{ij}=\begin{cases}
	\pm\frac{7D}{\sigma^2}((R/\sigma)^2-1)^6 (\vec{r}_i-\vec{r}_j) &R<\sigma\;,\\
	0 &R\geq \sigma\;.
\end{cases}
\]
We notice here that quite often, one introduces a repulsive force via a truncated Lennard-Jones potential (see \cite{Rex-Loewen-07,Pikovsky-21a,Aranson-Pikovsky-22}). However, to have a good performance of the Runge-Kutta fourth-order integration method, it is preferable to have maximally smooth potential; this property is ensured by choosing power $7$ in \eqref{eq:intpot}.

\begin{figure}[!htb]
	\centering
	\includegraphics[width=\columnwidth]{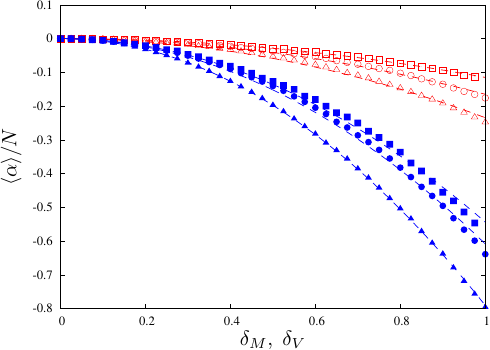}
	\caption{Particles with potential interaction (parameters $\sigma=1$, $D=10^4$, time of averaging $10^5$.)
		Red: rate vs $\delta_M$ for $\delta_V=0$ (the masses of particles are $1\pm \delta_M/2$); 
		blue: rate vs $\delta_V$ (the velocities of particles are $V=1\pm \delta_V/2$) for $\delta_M=0$. Squares: two particles, 
		circles: five particles; triangles: 10 particles. Lines: fits according to the square $\text{rate}\sim \delta^2$. All the rates are scaled by the particle number (i.e., convergence rate per particle).}
	\label{fig:phvol2}
\end{figure}

Calculations of the average divergence $\langle \alpha(t) \rangle_T$ for several interacting particles are presented in Fig.~\ref{fig:phvol2}. In these simulations, 
we either considered particles with equal masses $M=1$ and a uniform spread of velocities 
$V_m=1-\frac{1}{2}\delta_V,\ldots, 1+\frac{1}{2}\delta_V$;
or particles with  the same velocities $V=1$ and a uniform spread of masses $M_m=1-\frac{1}{2}\delta_M,\ldots, 1+\frac{1}{2}\delta_M$. One can see that for $\delta_M=\delta_V=0$, the phase volume is conserved, while the break of the particle's identity leads to convergence of this volume. The rate appears proportional to the square of the disorder levels $\sim \delta_M^2; \sim\delta_V^2$. 
The scaling with the number of particles $\langle \alpha(t) \rangle_T\sim N$ is only approximate,
because due to the fixed external potential, in a larger population, the particles are, on average, closer packed, and the interaction forces have a larger contribution to the phase volume changes. 

\section{Conclusion}
\label{sec:concl}
Summarizing, we have demonstrated that two popular models for
the deterministic active particles coincide in the overactive limit. In this limit, the particle's speed is constant (perfect ``cruise control''), and external forces govern the direction of the motion. These equations can be reformulated as a Hamiltonian system for motion in a static potential. The Hamilton dynamics is the same as in the ray optics. This is not surprising because light propagates with a constant speed and cannot be stopped but deviates due to the inhomogeneity of the refraction index. In this sense, overactive particles behave like photons. Because for two-dimensional motion, there are two degrees of freedom, one generally expects chaotic and regular domains in the phase space. We illustrated these types of dynamics for a particle in a harmonic potential; other confining potentials are expected to demonstrate similar features.  The overactive limit is based on the assumption that the internal active force is much stronger than external forces. For a harmonic potential, the external force is unlimited, and there indeed exist trajectories that climb toward large values of the potential where the external force becomes large. We show that such trajectories are strongly (in fact, faster than exponential) unstable, so in calculations, such events where the overactive limit could be violated practically do not appear.

An interesting feature of the overactive dynamics is that the Hamilton formulation appears to be valid for static external potentials only. Straightforward attempts to write a time-dependent Hamilton function in the case of a time-dependent potential and to formulate a many-body Hamilton function in the case of interacting particles failed. Thus, to characterize the conservativity of the dynamics, we focused on the conservation of the phase volume (which is ensured for Hamiltonian systems). We have shown analytically, that the phase volume is, on average, conserved for a particle in an external static potential and for interacting \textit{identical} particles. Performed numerical analysis demonstrated that the phase volume converges for a particle in a time-dependent potential and for interacting \textit{nonidentical} particles. An interesting question (a subject of ongoing research) is whether this non-conservativity can lead to observable effects like synchronization. 

Finally, we mention two generalizations of the presented approach to be reported elsewhere. First, we restricted the presentation to the 2-dimensional dynamics; the 3-dimensional case can be treated within the same framework. The second generalization includes the effect of chirality; in this case, a free overactive particle moves with a constant speed on a circular orbit. Preliminary calculations show that this motion in a static potential can also be formulated as a Hamiltonian system.

\acknowledgments
We thank I. Aranson, J. Brady, O. Dauchot, Ph. Nimphius, and F. Peruani for valuable discussions. 

\section*{Data availability}
All numerical simulations  are
described in the paper and can be reproduced without additional
information.
\bibliography{spp}

\begin{thebibliography}{10}%
\makeatletter
\providecommand \@ifxundefined [1]{%
 \ifx #1\undefined \expandafter \@firstoftwo
 \else \expandafter \@secondoftwo
\fi
}%
\providecommand \@ifnum [1]{%
 \ifnum #1\expandafter \@firstoftwo
 \else \expandafter \@secondoftwo
\fi
}%
\providecommand \enquote [1]{``#1''}%
\providecommand \bibnamefont  [1]{#1}%
\providecommand \bibfnamefont [1]{#1}%
\providecommand \citenamefont [1]{#1}%
\providecommand\href[0]{\@sanitize\@href}%
\providecommand\@href[1]{\endgroup\@@startlink{#1}\endgroup\@@href}%
\providecommand\@@href[1]{#1\@@endlink}%
\providecommand \@sanitize [0]{\begingroup\catcode`\&12\catcode`\#12\relax}%
\@ifxundefined \pdfoutput {\@firstoftwo}{%
 \@ifnum{\z@=\pdfoutput}{\@firstoftwo}{\@secondoftwo}%
}{%
 \providecommand\@@startlink[1]{\leavevmode\special{html:<a href="#1">}}%
 \providecommand\@@endlink[0]{\special{html:</a>}}%
}{%
 \providecommand\@@startlink[1]{%
  \leavevmode
  \pdfstartlink
   attr{/Border[0 0 1 ]/H/I/C[0 1 1]}%
   user{/Subtype/Link/A<</Type/Action/S/URI/URI(#1)>>}%
  \relax
 }%
 \providecommand\@@endlink[0]{\pdfendlink}%
}%
\providecommand \url  [0]{\begingroup\@sanitize \@url }%
\providecommand \@url [1]{\endgroup\@href {#1}{\urlprefix}}%
\providecommand \urlprefix [0]{URL }%
\providecommand \Eprint[0]{\href }%
\@ifxundefined \urlstyle {%
  \providecommand \doi [1]{doi:\discretionary{}{}{}#1}%
}{%
  \providecommand \doi [0]{doi:\discretionary{}{}{}\begingroup
  \urlstyle{rm}\Url }%
}%
\providecommand \doibase [0]{http://dx.doi.org/}%
\providecommand \Doi[1]{\href{\doibase#1}}%
\providecommand \selectlanguage [0]{\@gobble}%
\providecommand \bibinfo [0]{\@secondoftwo}%
\providecommand \bibfield [0]{\@secondoftwo}%
\providecommand \translation [1]{[#1]}%
\providecommand \BibitemOpen[0]{}%
\providecommand \bibitemStop [0]{}%
\providecommand \bibitemNoStop [0]{.\EOS\space}%
\providecommand \EOS [0]{\spacefactor3000\relax}%
\providecommand \BibitemShut [1]{\csname bibitem#1\endcsname}%
\bibitem{chate2020dry}%
  \BibitemOpen
  \bibfield{author}{%
  \bibinfo {author} {\bibfnamefont{Hugues}\ \bibnamefont{Chat{\'e}}},\ }%
  \bibfield{title}{%
  \enquote{\bibinfo {title} {Dry aligning dilute active matter},}\ }%
  \bibfield{journal}{%
  \bibinfo {journal} {Annual Review of Condensed Matter Physics}\ }%
  \textbf{\bibinfo {volume} {11}},\ \bibinfo {pages} {189--212} (\bibinfo
  {year} {2020})\BibitemShut{NoStop}%
\bibitem{gompper20202020}%
  \BibitemOpen
  \bibfield{author}{%
  \bibinfo {author} {\bibfnamefont{Gerhard}\ \bibnamefont{Gompper}}, \bibinfo
  {author} {\bibfnamefont{Roland~G}\ \bibnamefont{Winkler}}, \bibinfo {author}
  {\bibfnamefont{Thomas}\ \bibnamefont{Speck}}, \bibinfo {author}
  {\bibfnamefont{Alexandre}\ \bibnamefont{Solon}}, \bibinfo {author}
  {\bibfnamefont{Cesare}\ \bibnamefont{Nardini}}, \bibinfo {author}
  {\bibfnamefont{Fernando}\ \bibnamefont{Peruani}}, \bibinfo {author}
  {\bibfnamefont{Hartmut}\ \bibnamefont{L{\"o}wen}}, \bibinfo {author}
  {\bibfnamefont{Ramin}\ \bibnamefont{Golestanian}}, \bibinfo {author}
  {\bibfnamefont{U~Benjamin}\ \bibnamefont{Kaupp}}, \bibinfo {author}
  {\bibfnamefont{Luis}\ \bibnamefont{Alvarez}}, \emph{et~al.},\ }%
  \bibfield{title}{%
  \enquote{\bibinfo {title} {The 2020 motile active matter roadmap},}\ }%
  \bibfield{journal}{%
  \bibinfo {journal} {Journal of Physics: Condensed Matter}\ }%
  \textbf{\bibinfo {volume} {32}},\ \bibinfo {pages} {193001} (\bibinfo {year}
  {2020})\BibitemShut{NoStop}%
\bibitem{aranson2013active}%
  \BibitemOpen
  \bibfield{author}{%
  \bibinfo {author} {\bibfnamefont{Igor~S}\ \bibnamefont{Aranson}},\ }%
  \bibfield{title}{%
  \enquote{\bibinfo {title} {Active colloids},}\ }%
  \bibfield{journal}{%
  \bibinfo {journal} {Physics-Uspekhi}\ }%
  \textbf{\bibinfo {volume} {56}},\ \bibinfo {pages} {79} (\bibinfo {year}
  {2013})\BibitemShut{NoStop}%
\bibitem{bechinger2016active}%
  \BibitemOpen
  \bibfield{author}{%
  \bibinfo {author} {\bibfnamefont{Clemens}\ \bibnamefont{Bechinger}}, \bibinfo
  {author} {\bibfnamefont{Roberto}\ \bibnamefont{Di~Leonardo}}, \bibinfo
  {author} {\bibfnamefont{Hartmut}\ \bibnamefont{L{\"o}wen}}, \bibinfo {author}
  {\bibfnamefont{Charles}\ \bibnamefont{Reichhardt}}, \bibinfo {author}
  {\bibfnamefont{Giorgio}\ \bibnamefont{Volpe}},\ and\ \bibinfo {author}
  {\bibfnamefont{Giovanni}\ \bibnamefont{Volpe}},\ }%
  \bibfield{title}{%
  \enquote{\bibinfo {title} {Active particles in complex and crowded
  environments},}\ }%
  \bibfield{journal}{%
  \bibinfo {journal} {Reviews of Modern Physics}\ }%
  \textbf{\bibinfo {volume} {88}},\ \bibinfo {pages} {045006} (\bibinfo {year}
  {2016})\BibitemShut{NoStop}%
\bibitem{peruani2018cold}%
  \BibitemOpen
  \bibfield{author}{%
  \bibinfo {author} {\bibfnamefont{Fernando}\ \bibnamefont{Peruani}}\ and\
  \bibinfo {author} {\bibfnamefont{Igor~S}\ \bibnamefont{Aranson}},\ }%
  \bibfield{title}{%
  \enquote{\bibinfo {title} {Cold active motion: how time-independent disorder
  affects the motion of self-propelled agents},}\ }%
  \bibfield{journal}{%
  \bibinfo {journal} {Physical Review Letters}\ }%
  \textbf{\bibinfo {volume} {120}},\ \bibinfo {pages} {238101} (\bibinfo {year}
  {2018})\BibitemShut{NoStop}%
\bibitem{erdmann2000brownian}%
  \BibitemOpen
  \bibfield{author}{%
  \bibinfo {author} {\bibfnamefont{Udo}\ \bibnamefont{Erdmann}}, \bibinfo
  {author} {\bibfnamefont{Werner}\ \bibnamefont{Ebeling}}, \bibinfo {author}
  {\bibfnamefont{Lutz}\ \bibnamefont{Schimansky-Geier}},\ and\ \bibinfo
  {author} {\bibfnamefont{Frank}\ \bibnamefont{Schweitzer}},\ }%
  \bibfield{title}{%
  \enquote{\bibinfo {title} {Brownian particles far from equilibrium},}\ }%
  \bibfield{journal}{%
  \bibinfo {journal} {The European Physical Journal B-Condensed Matter and
  Complex Systems}\ }%
  \textbf{\bibinfo {volume} {15}},\ \bibinfo {pages} {105--113} (\bibinfo
  {year} {2000})\BibitemShut{NoStop}%
\bibitem{romanczuk2012active}%
  \BibitemOpen
  \bibfield{author}{%
  \bibinfo {author} {\bibfnamefont{Pawel}\ \bibnamefont{Romanczuk}}, \bibinfo
  {author} {\bibfnamefont{Markus}\ \bibnamefont{B{\"a}r}}, \bibinfo {author}
  {\bibfnamefont{Werner}\ \bibnamefont{Ebeling}}, \bibinfo {author}
  {\bibfnamefont{Benjamin}\ \bibnamefont{Lindner}},\ and\ \bibinfo {author}
  {\bibfnamefont{Lutz}\ \bibnamefont{Schimansky-Geier}},\ }%
  \bibfield{title}{%
  \enquote{\bibinfo {title} {Active brownian particles: From individual to
  collective stochastic dynamics},}\ }%
  \bibfield{journal}{%
  \bibinfo {journal} {The European Physical Journal Special Topics}\ }%
  \textbf{\bibinfo {volume} {202}},\ \bibinfo {pages} {1--162} (\bibinfo {year}
  {2012})\BibitemShut{NoStop}%
\bibitem{Aranson-Pikovsky-22}%
  \BibitemOpen
  \bibfield{author}{%
  \bibinfo {author} {\bibfnamefont{Igor~S.}\ \bibnamefont{Aranson}}\ and\
  \bibinfo {author} {\bibfnamefont{Arkady}\ \bibnamefont{Pikovsky}},\ }%
  \bibfield{title}{%
  \enquote{\bibinfo {title} {Confinement and collective escape of active
  particles},}\ }%
  \bibfield{journal}{%
  \Doi{10.1103/PhysRevLett.128.108001}{\bibinfo {journal} {Phys. Rev. Lett.}}\
  }%
  \textbf{\bibinfo {volume} {128}},\ \bibinfo {pages} {108001} (\bibinfo
  {month} {Mar}\ \bibinfo {year} {2022}),\
  \url{https://link.aps.org/doi/10.1103/PhysRevLett.128.108001}\BibitemShut{NoStop}%
\bibitem{erdmann2002excitation}%
  \BibitemOpen
  \bibfield{author}{%
  \bibinfo {author} {\bibfnamefont{Udo}\ \bibnamefont{Erdmann}}, \bibinfo
  {author} {\bibfnamefont{Werner}\ \bibnamefont{Ebeling}},\ and\ \bibinfo
  {author} {\bibfnamefont{Vadim~S}\ \bibnamefont{Anishchenko}},\ }%
  \bibfield{title}{%
  \enquote{\bibinfo {title} {Excitation of rotational modes in two-dimensional
  systems of driven brownian particles},}\ }%
  \bibfield{journal}{%
  \bibinfo {journal} {Physical Review E}\ }%
  \textbf{\bibinfo {volume} {65}},\ \bibinfo {pages} {061106} (\bibinfo {year}
  {2002})\BibitemShut{NoStop}%
\bibitem{erdmann2005attractors}%
  \BibitemOpen
  \bibfield{author}{%
  \bibinfo {author} {\bibfnamefont{Udo}\ \bibnamefont{Erdmann}}\ and\ \bibinfo
  {author} {\bibfnamefont{Werner}\ \bibnamefont{Ebeling}},\ }%
  \bibfield{title}{%
  \enquote{\bibinfo {title} {On the attractors of two-dimensional rayleigh
  oscillators including noise},}\ }%
  \bibfield{journal}{%
  \bibinfo {journal} {International Journal of Bifurcation and Chaos}\ }%
  \textbf{\bibinfo {volume} {15}},\ \bibinfo {pages} {3623--3633} (\bibinfo
  {year} {2005})\BibitemShut{NoStop}%
\bibitem{Lam_2015}%
  \BibitemOpen
  \bibfield{author}{%
  \bibinfo {author} {\bibfnamefont{Khanh-Dang Nguyen~Thu}\ \bibnamefont{Lam}},
  \bibinfo {author} {\bibfnamefont{Michael}\ \bibnamefont{Schindler}},\ and\
  \bibinfo {author} {\bibfnamefont{Olivier}\ \bibnamefont{Dauchot}},\ }%
  \bibfield{title}{%
  \enquote{\bibinfo {title} {Self-propelled hard disks: implicit alignment and
  transition to collective motion},}\ }%
  \bibfield{journal}{%
  \Doi{10.1088/1367-2630/17/11/113056}{\bibinfo {journal} {New Journal of
  Physics}}\ }%
  \textbf{\bibinfo {volume} {17}},\ \bibinfo {pages} {113056} (\bibinfo {month}
  {nov}\ \bibinfo {year} {2015}),\
  \url{https://dx.doi.org/10.1088/1367-2630/17/11/113056}\BibitemShut{NoStop}%
\bibitem{PhysRevLett.122.068002}%
  \BibitemOpen
  \bibfield{author}{%
  \bibinfo {author} {\bibfnamefont{Olivier}\ \bibnamefont{Dauchot}}\ and\
  \bibinfo {author} {\bibfnamefont{Vincent}\ \bibnamefont{D\'emery}},\ }%
  \bibfield{title}{%
  \enquote{\bibinfo {title} {Dynamics of a self-propelled particle in a
  harmonic trap},}\ }%
  \bibfield{journal}{%
  \Doi{10.1103/PhysRevLett.122.068002}{\bibinfo {journal} {Phys. Rev. Lett.}}\
  }%
  \textbf{\bibinfo {volume} {122}},\ \bibinfo {pages} {068002} (\bibinfo
  {month} {Feb}\ \bibinfo {year} {2019}),\
  \url{https://link.aps.org/doi/10.1103/PhysRevLett.122.068002}\BibitemShut{NoStop}%
\bibitem{baconnier2022selective}%
  \BibitemOpen
  \bibfield{author}{%
  \bibinfo {author} {\bibfnamefont{Paul}\ \bibnamefont{Baconnier}}, \bibinfo
  {author} {\bibfnamefont{Dor}\ \bibnamefont{Shohat}}, \bibinfo {author}
  {\bibfnamefont{C~Hern{\'a}ndez}\ \bibnamefont{L{\'o}pez}}, \bibinfo {author}
  {\bibfnamefont{Corentin}\ \bibnamefont{Coulais}}, \bibinfo {author}
  {\bibfnamefont{Vincent}\ \bibnamefont{D{\'e}mery}}, \bibinfo {author}
  {\bibfnamefont{Gustavo}\ \bibnamefont{D{\"u}ring}},\ and\ \bibinfo {author}
  {\bibfnamefont{Olivier}\ \bibnamefont{Dauchot}},\ }%
  \bibfield{title}{%
  \enquote{\bibinfo {title} {Selective and collective actuation in active
  solids},}\ }%
  \bibfield{journal}{%
  \bibinfo {journal} {Nature Physics}\ }%
  \textbf{\bibinfo {volume} {18}},\ \bibinfo {pages} {1234--1239} (\bibinfo
  {year} {2022})\BibitemShut{NoStop}%
\bibitem{Damascena_etal-22}%
  \BibitemOpen
  \bibfield{author}{%
  \bibinfo {author} {\bibfnamefont{Rubens~H.}\ \bibnamefont{Damascena}},
  \bibinfo {author} {\bibfnamefont{Leonardo R.~E.}\ \bibnamefont{Cabral}},\
  and\ \bibinfo {author} {\bibfnamefont{Cl\'ecio C. de~Souza}\
  \bibnamefont{Silva}},\ }%
  \bibfield{title}{%
  \enquote{\bibinfo {title} {Coexisting orbits and chaotic dynamics of a
  confined self-propelled particle},}\ }%
  \bibfield{journal}{%
  \Doi{10.1103/PhysRevE.105.064608}{\bibinfo {journal} {Phys. Rev. E}}\ }%
  \textbf{\bibinfo {volume} {105}},\ \bibinfo {pages} {064608} (\bibinfo
  {month} {Jun}\ \bibinfo {year} {2022}),\
  \url{https://link.aps.org/doi/10.1103/PhysRevE.105.064608}\BibitemShut{NoStop}%
\bibitem{Kravtsov-Orlov-90}%
  \BibitemOpen
  \bibfield{author}{%
  \bibinfo {author} {\bibfnamefont{Yu.~A.}\ \bibnamefont{Kravtsov}}\ and\
  \bibinfo {author} {\bibfnamefont{Yu.~I.}\ \bibnamefont{Orlov}},\ }%
  \emph{\bibinfo {title} {Geometrical Optics of Inhomogeneous Media}}\
  (\bibinfo {publisher} {Springer},\ \bibinfo {address} {Berlin, Heidelberg},\
  \bibinfo {year} {1990})\BibitemShut{NoStop}%
\bibitem{Ross_etal-22}%
  \BibitemOpen
  \bibfield{author}{%
  \bibinfo {author} {\bibfnamefont{Tyler~D.}\ \bibnamefont{Ross}}, \bibinfo
  {author} {\bibfnamefont{Dino}\ \bibnamefont{Osmanović}}, \bibinfo {author}
  {\bibfnamefont{John~F.}\ \bibnamefont{Brady}},\ and\ \bibinfo {author}
  {\bibfnamefont{Paul W.~K.}\ \bibnamefont{Rothemund}},\ }%
  \bibfield{title}{%
  \enquote{\bibinfo {title} {Ray optics for gliders},}\ }%
  \bibfield{journal}{%
  \bibinfo {journal} {ACS Nano}\ }%
  \textbf{\bibinfo {volume} {16}},\ \bibinfo {pages} {16191--16200} (\bibinfo
  {year} {2022})\BibitemShut{NoStop}%
\bibitem{chepizhko2013diffusion}%
  \BibitemOpen
  \bibfield{author}{%
  \bibinfo {author} {\bibfnamefont{Oleksandr}\ \bibnamefont{Chepizhko}}\ and\
  \bibinfo {author} {\bibfnamefont{Fernando}\ \bibnamefont{Peruani}},\ }%
  \bibfield{title}{%
  \enquote{\bibinfo {title} {Diffusion, subdiffusion, and trapping of active
  particles in heterogeneous media},}\ }%
  \bibfield{journal}{%
  \bibinfo {journal} {Physical review letters}\ }%
  \textbf{\bibinfo {volume} {111}},\ \bibinfo {pages} {160604} (\bibinfo {year}
  {2013})\BibitemShut{NoStop}%
\bibitem{touma2010self}%
  \BibitemOpen
  \bibfield{author}{%
  \bibinfo {author} {\bibfnamefont{Jihad~R}\ \bibnamefont{Touma}}, \bibinfo
  {author} {\bibfnamefont{Amer}\ \bibnamefont{Shreim}},\ and\ \bibinfo {author}
  {\bibfnamefont{Leonid~I}\ \bibnamefont{Klushin}},\ }%
  \bibfield{title}{%
  \enquote{\bibinfo {title} {Self-organization in two-dimensional swarms},}\ }%
  \bibfield{journal}{%
  \bibinfo {journal} {Physical Review E}\ }%
  \textbf{\bibinfo {volume} {81}},\ \bibinfo {pages} {066106} (\bibinfo {year}
  {2010})\BibitemShut{NoStop}%
\bibitem{Alsayegh_etal-16}%
  \BibitemOpen
  \bibfield{author}{%
  \bibinfo {author} {\bibfnamefont{Amara~A.}\ \bibnamefont{Al~Sayegh}},
  \bibinfo {author} {\bibfnamefont{Leonid}\ \bibnamefont{Klushin}},\ and\
  \bibinfo {author} {\bibfnamefont{Jihad}\ \bibnamefont{Touma}},\ }%
  \bibfield{title}{%
  \enquote{\bibinfo {title} {Steady and transient states in low-energy swarms:
  Stability and first-passage times},}\ }%
  \bibfield{journal}{%
  \Doi{10.1103/PhysRevE.93.032602}{\bibinfo {journal} {Phys. Rev. E}}\ }%
  \textbf{\bibinfo {volume} {93}},\ \bibinfo {pages} {032602} (\bibinfo {month}
  {Mar}\ \bibinfo {year} {2016}),\
  \url{https://link.aps.org/doi/10.1103/PhysRevE.93.032602}\BibitemShut{NoStop}%
\bibitem{PhysRevLett.120.208001}%
  \BibitemOpen
  \bibfield{author}{%
  \bibinfo {author} {\bibfnamefont{Guillaume}\ \bibnamefont{Briand}}, \bibinfo
  {author} {\bibfnamefont{Michael}\ \bibnamefont{Schindler}},\ and\ \bibinfo
  {author} {\bibfnamefont{Olivier}\ \bibnamefont{Dauchot}},\ }%
  \bibfield{title}{%
  \enquote{\bibinfo {title} {Spontaneously flowing crystal of self-propelled
  particles},}\ }%
  \bibfield{journal}{%
  \Doi{10.1103/PhysRevLett.120.208001}{\bibinfo {journal} {Phys. Rev. Lett.}}\
  }%
  \textbf{\bibinfo {volume} {120}},\ \bibinfo {pages} {208001} (\bibinfo
  {month} {May}\ \bibinfo {year} {2018}),\
  \url{https://link.aps.org/doi/10.1103/PhysRevLett.120.208001}\BibitemShut{NoStop}%
\bibitem{Rex-Loewen-07}%
  \BibitemOpen
  \bibfield{author}{%
  \bibinfo {author} {\bibfnamefont{M.}~\bibnamefont{Rex}}\ and\ \bibinfo
  {author} {\bibfnamefont{H.}~\bibnamefont{L\"owen}},\ }%
  \bibfield{title}{%
  \enquote{\bibinfo {title} {Lane formation in oppositely charged colloids
  driven by an electric field: Chaining and two-dimensional crystallization},}\
  }%
  \bibfield{journal}{%
  \bibinfo {journal} {Phys. Rev. E}\ }%
  \textbf{\bibinfo {volume} {75}},\ \bibinfo {pages} {051402} (\bibinfo {year}
  {2007})\BibitemShut{NoStop}%
\bibitem{Pikovsky-21a}%
  \BibitemOpen
  \bibfield{author}{%
  \bibinfo {author} {\bibfnamefont{Arkady}\ \bibnamefont{Pikovsky}},\ }%
  \bibfield{title}{%
  \enquote{\bibinfo {title} {Transition to synchrony in chiral active
  particles},}\ }%
  \bibfield{journal}{%
  \bibinfo {journal} {J. Phys. Complexity}\ }%
  \textbf{\bibinfo {volume} {2}},\ \bibinfo {pages} {025009} (\bibinfo {year}
  {2021})\BibitemShut{NoStop}%
\end{thebibliography}%

\end{document}